\documentclass[prl,twocolumn,superscriptaddress,nofootinbib]{revtex4-1}

\usepackage{comment}
\usepackage{slashed}
\usepackage{latexsym}
\usepackage{graphics}
\usepackage{bm}
\usepackage{color}
\usepackage{amsmath}
\usepackage{amssymb}
\usepackage{mathrsfs}
\usepackage{graphicx}
\usepackage{slashed}
\usepackage{siunitx}
\usepackage{braket}
\usepackage{calligra}

\definecolor{linkcolor}{rgb}{0.7752941176470588, 0.22078431372549023, 0.2262745098039215}

\usepackage{hyperref}
\hypersetup{colorlinks=true,
linkcolor=linkcolor,
citecolor=linkcolor,
urlcolor=linkcolor,
linktocpage=true
}

\def\lsim{\mathrel{\rlap{\lower4pt\hbox{\hskip1pt$\sim$}}
    \raise1pt\hbox{$<$}}}                
\def\gsim{\mathrel{\rlap{\lower4pt\hbox{\hskip1pt$\sim$}}
    \raise1pt\hbox{$>$}}}                
\def\dd{\textrm{d}}
\newcommand{\x}{\chi}
\newcommand{\p}{\prime}
\newcommand{\mAp}{m_{A^\prime}}
\newcommand{\Ap}{A^\prime}
\newcommand{\eps}{\epsilon}
\newcommand{\qeff}{q_\text{eff}}
\newcommand{\rsc}{\hspace{-0.06cm} \text{\LARGE \textcalligra{r}}}

\definecolor{nicegreen}{rgb}{0.1,0.5,0.1}

\DeclareSIUnit\electronvolt{e\kern-.05em V}
\DeclareSIUnit\tonneyear{tonne-year}

\makeatletter 
    
\renewcommand\onecolumngrid{
\do@columngrid{one}{\@ne}%
\def\set@footnotewidth{\onecolumngrid}
\def\footnoterule{\kern-6pt\hrule width 1.5in\kern6pt}%
}

\renewcommand\twocolumngrid{
        \def\footnoterule{
        \dimen@\skip\footins\divide\dimen@\thr@@
        \kern-\dimen@\hrule width.5in\kern\dimen@}
        \do@columngrid{mlt}{\tw@}
}%

\makeatother 

\begin{document}

\title{Low-Energy Signals from the Formation of Dark Matter-Nuclear Bound States}

\author{Asher Berlin}\email{ajb643@nyu.edu}
\affiliation{Center for Cosmology and Particle Physics, Department of Physics, New York University, New York, NY 10003, USA}

\author{Hongwan Liu}\email{hongwanl@princeton.edu}
\affiliation{Center for Cosmology and Particle Physics, Department of Physics, New York University, New York, NY 10003, USA}
\affiliation{Department of Physics, Princeton University, Princeton, New Jersey, 08544, USA}

\author{Maxim Pospelov}\email{pospelov@umn.edu}
\affiliation{School of Physics and Astronomy, University of Minnesota, Minneapolis, MN 55455, USA}
\affiliation{William I. Fine Theoretical Physics Institute, School of Physics and Astronomy,
University of Minnesota, Minneapolis, MN 55455, USA}

\author{Harikrishnan Ramani}\email{hramani@stanford.edu}
\affiliation{Stanford Institute for Theoretical Physics,
Stanford University, Stanford, CA 94305, USA}

\begin{abstract}
Dark matter particles may bind with nuclei if there exists an attractive force of sufficient strength. We show that a dark photon mediator of mass $\sim (10 - 100) \ \text{MeV}$ that kinetically mixes with Standard Model electromagnetism at the level of $\sim 10^{-3}$ generates keV-scale binding energies between dark matter and heavy elements, while forbidding the ability to bind with light elements. In underground direct detection experiments, the formation of such bound states liberates keV-scale energy in the form of electrons and photons, giving rise to mono-energetic electronic signals with a time-structure that may contain daily and seasonal modulations. We show that data from liquid-xenon detectors provides exquisite sensitivity to this scenario, constraining the galactic abundance of such dark  particles to be at most $\sim 10^{-18} - 10^{-12}$ of the galactic dark matter density for masses spanning $\sim (1 - 10^5) \ \text{GeV}$. However, an exponentially small fractional abundance of these dark particles is enough to explain the observed electron recoil excess at XENON1T.

\end{abstract}

\maketitle

\emph{Introduction.} Over the years, dark matter (DM) direct detection experiments have become extraordinarily sensitive to sub-MeV energy deposition by exotic sources, with thresholds recently extending down to sub-keV energies. While the primary motivation for these experiments is to search for the elastic scattering of weakly interacting massive particles (WIMPs) off nuclei, the scope of these searches now includes, e.g., electron scattering, DM absorption, and exo- and endo-thermic inelasticity in DM-nucleus scattering (see, e.g., Ref.~\cite{Lin:2019uvt} for a review). Each of these searches seeks to measure the energy deposited from either the kinetic (scattering) or mass energy (absorption, inelasticity) of the incoming DM particle. In this paper, we point out a third, distinct alternative: the energy released due to the formation of a bound state between DM and a Standard Model (SM) nucleus. 

The phenomenology of DM-nuclear bound state formation has been studied previously in the literature. For instance, \SI{}{\mega\eV}-scale DM-nuclear binding was investigated in Refs.~\cite{Pospelov:2007xh,An:2012bs,Fornal:2020bzz}, while Refs.~\cite{Wallemacq:2013hsa, Wallemacq:2014lba, Wallemacq:2014sta, Laletin:2019qca} focused on \SI{}{\keV}-scale bound states involving dark atoms to explain the DAMA anomaly~\cite{DAMA:2008jlt,DAMA:2010gpn,Bernabei:2018jrt}. In contrast to these previously studied models, all of which require some degree of intricate model-building, we instead consider bound states arising from one of the simplest and most studied models in the literature within the last fifteen years~\cite{Boehm:2003hm,ArkaniHamed:2008qn,Pospelov:2007mp}: a DM particle $\x$ charged under a massive kinetically-mixed dark photon. We focus on a scenario where $\x$ possesses a sizeable interaction with normal matter and constitutes a small fraction $f_\x \ll 1$ of the total galactic DM density. This model naturally leads to \emph{i)} \SI{}{\keV}-scale DM-nuclear binding energies $E_B$ and \emph{ii)} preferential binding  to heavy nuclei, such that upon penetrating the terrestrial overburden, $\x$ only binds with the much heavier nuclei (such as xenon and thallium) commonly found in underground DM detectors. 

In a direct detection experiment whose target material consists of atoms $A$ of sufficiently large atomic number and mass, the process of DM-atom ``recombination'' $\x + A \to (\x A) + E_B$ releases electromagnetic energy $E_B$ equal to the binding energy of $(\x A)$.\footnote{We adopt the notation where a bound state is denoted by parentheses surrounding the names of the constituent particles.} For this reaction to occur, the minimum required coupling between $\x$ and $A$ is sufficiently large such that the galactic $\x$ population quickly thermalizes upon encountering Earth's environment, cooling down to terrestrial temperatures. The implications of thermalizing with the terrestrial environment are two-fold. First, the terrestrial $\x$ density is drastically enhanced compared to the galactic population, due to conservation of flux (the ``traffic-jam" scenario discussed in Refs.~\cite{Pospelov:2019vuf,Pospelov:2020ktu,Neufeld:2018slx}). 
Second, there are no observable elastic scattering signals of $\x$ despite its large couplings and enhanced terrestrial density, since the thermal energy of underground laboratory environments ($\lesssim \SI{300}{\kelvin} \sim \SI{25}{\milli\eV}$) is well below existing kinematic thresholds. However, since $E_B$ naturally lies near the keV-scale, the formation of DM-nuclear bound states is readily detectable at large scale experiments designed to search for WIMP-nuclear scattering. 

Among such direct detection experiments, the suite of large scale dual-phase xenon detectors plays an especially important role. For instance, ionization-only data from XENON10 and XENON100 place some of the strongest constraints on MeV-scale DM-electron scattering~\cite{Essig:2017kqs}, and the large exposure and low background counts (below $10^{-5} \text{ per kg-day-keV}$) of the XENON1T experiment enable new benchmark sensitivity not only to WIMP-nuclear scattering but also to sub-keV electronic recoils~\cite{Aprile:2018dbl}. Intriguingly, the XENON1T collaboration recently reported an excess of events consistent with electron recoils with an energy deposition of $(2 - 3) \ \text{keV}$~\cite{Aprile:2020tmw}. This may be consistent with a variety of recently proposed new physics models, all of which invoke a substantial flux of particles that feebly interact with normal matter (e.g., dark photons~\cite{Alonso-Alvarez:2020cdv,An:2020bxd}, neutrinos and dark radiation~\cite{Bloch:2020uzh,McKeen:2020vpf,Farzan:2020dds,Brdar:2020quo,Kuo:2021mtp}, and exothermic DM~\cite{Baryakhtar:2020rwy,Bramante:2020zos,Bloch:2020uzh,Harigaya:2020ckz}). In this Letter, we find that the observed rate at XENON1T may be explained as a result of a strongly-coupled particle that makes up an extremely small fraction $f_\x$ of the galactic DM density. In particular, we find that a DM subcomponent that binds to xenon nuclei with $E_B = \SI{2.5}{\kilo\eV}$ is a viable explanation to this anomaly. More generally, these signals are significantly constrained by XENON1T for fractional abundances greater than $10^{-18} \lesssim f_\x \lesssim 10^{-12}$ and particle masses spanning $\SI{1}{\giga\eV} \lesssim m_\x \lesssim \SI{30}{\tera\eV}$.

\emph{Model and Bound State Parameter Space.} We consider a subcomponent of DM $\x$ that is directly charged under a new massive dark $U(1)$ gauge boson $A^\p_\mu$ that kinetically mixes with the SM photon,\footnote{For concreteness, our calculations assume that $\x$ is fermionic, although all considerations in this paper apply equally well to scalar DM.} 
\begin{equation}
\mathcal{L} \supset - \frac{\epsilon}{2} \, F_{\mu\nu}^\prime  \, F^{\mu\nu} + \frac{\mAp^2}{2} \, A_\mu^{\p \, 2}
~,
\end{equation}
where $\mAp$ is the dark photon mass and $\eps \ll 1$ controls the strength of kinetic mixing~\cite{Holdom:1985ag}. If $\eps$ is generated radiatively from particles charged under both the SM and dark sector, the natural expectation is $\eps \sim (\alpha_D \, \alpha)^{1/2} / 4 \pi$, where $\alpha_D$ and $\alpha$ are the dark photon and SM fine-structure constant, respectively. $\x$ interacts with normal matter through a small effective coupling $e \qeff$ where $\qeff \equiv \eps \sqrt{\alpha_D / \alpha}$. 

The dark photon also mediates attractive self-interactions, such that resonances and capture to $(\x \bar{\x})$ bound states can significantly reduce the cosmological $\x$ density~\cite{ArkaniHamed:2008qn,Pospelov:2008jd,Feng:2010zp,Cirelli:2016rnw,An:2016gad}. It is therefore reasonable to consider a small fraction of the DM energy density $f_\x \equiv \rho_\x / \rho_{_\text{DM}}$ that is composed of such particles. In the local vicinity of the galaxy, we consider $f_\x$ to be a free parameter, noting that deviations from a standard thermal cosmological history could result in $f_\x \ll 1$.\footnote{As a concrete example, arbitrarily small abundances of $\x$ are cosmologically generated provided that the reheat temperature of the universe $T_\text{RH}$ is significantly smaller than $m_\x$. In this case, electron annihilations freeze-in a fractional abundance of $\x$ corresponding to $f_\x \sim (\alpha \qeff)^2 e^{-2 m_\x / T_\text{RH}} m_\x \, m_\text{pl}/ (T_\text{RH} \, T_\text{mre})$, where $m_\text{pl} \sim 10^{19} \ \text{GeV}$ is the Planck mass and $T_\text{mre}\sim 1 \ \text{eV}$ is the temperature at matter-radiation equality.}

\begin{figure}
\hspace*{-0.2in}
\vspace*{0.5cm}
\includegraphics[scale=0.70]{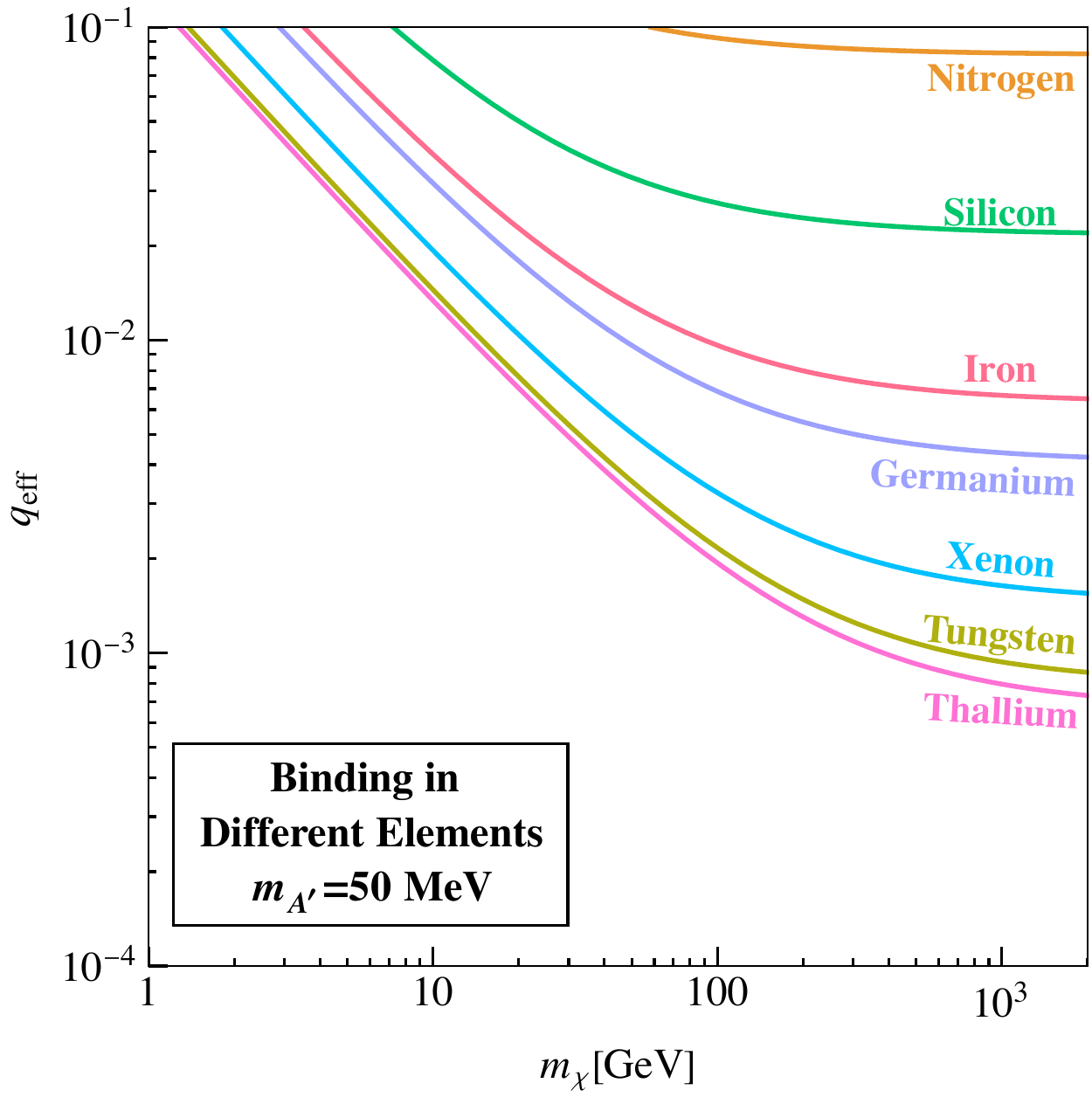}
\caption{The minimum coupling $\qeff$ required for dark matter to bind with various nuclei, as a function of dark matter mass $m_\x$, fixing $\mAp = 50 \ \text{MeV}$. A dark matter-nuclear bound state exists with an element for values of $\qeff$ above the corresponding line.} 
\label{fig:stability}
\end{figure}

A massive dark photon only allows binding with heavy nuclei. Intuitively, this selection arises because the characteristic size of the bound state $(\qeff Z \alpha \mu)^{-1}$ should be smaller than the range of the interaction $\mAp^{-1}$, with both the atomic number $Z$ of the nucleus and the DM-nuclear reduced mass $\mu$ increasing with larger nuclei. To incorporate effects associated with the finite size of the nucleus, we use Bargmann's limit~\cite{Bargmann961}, which can be used to show that a DM-nuclear bound state exists only if
\begin{equation}
\label{Bargmann1}
\qeff Z \alpha \, \mu \gtrsim \mAp \times 
\begin{cases}
1/2 &(\mAp R_\text{nuc} \ll 1)
\\
\mAp R_\text{nuc} / 3 &(\mAp R_\text{nuc} \gg 1)
~,
\end{cases}
\end{equation}
where the radius of a nucleus of atomic mass $A$ is $R_\text{nuc} \sim \SI{1.1}{\femto\meter} \times A^{1/3}$. In Fig.~\ref{fig:stability}, we show the minimum value of $\qeff$ that is required for $\x$ to bind with various nuclei as a function of the DM mass $m_\x$, fixing $\mAp = \SI{50}{\mega\eV}$. We see a clear preference to bind to heavier elements for all $\x$ masses. This is most pronounced for large DM masses $m_\x \gg \SI{100}{\giga\eV}$, in which case $\mu \simeq m_N$ and the minimal coupling to bind strongly depends on the nuclear mass. 

\begin{figure}
\hspace*{-0.1in}
\vspace*{0.1cm}
\includegraphics[width=0.47\textwidth]{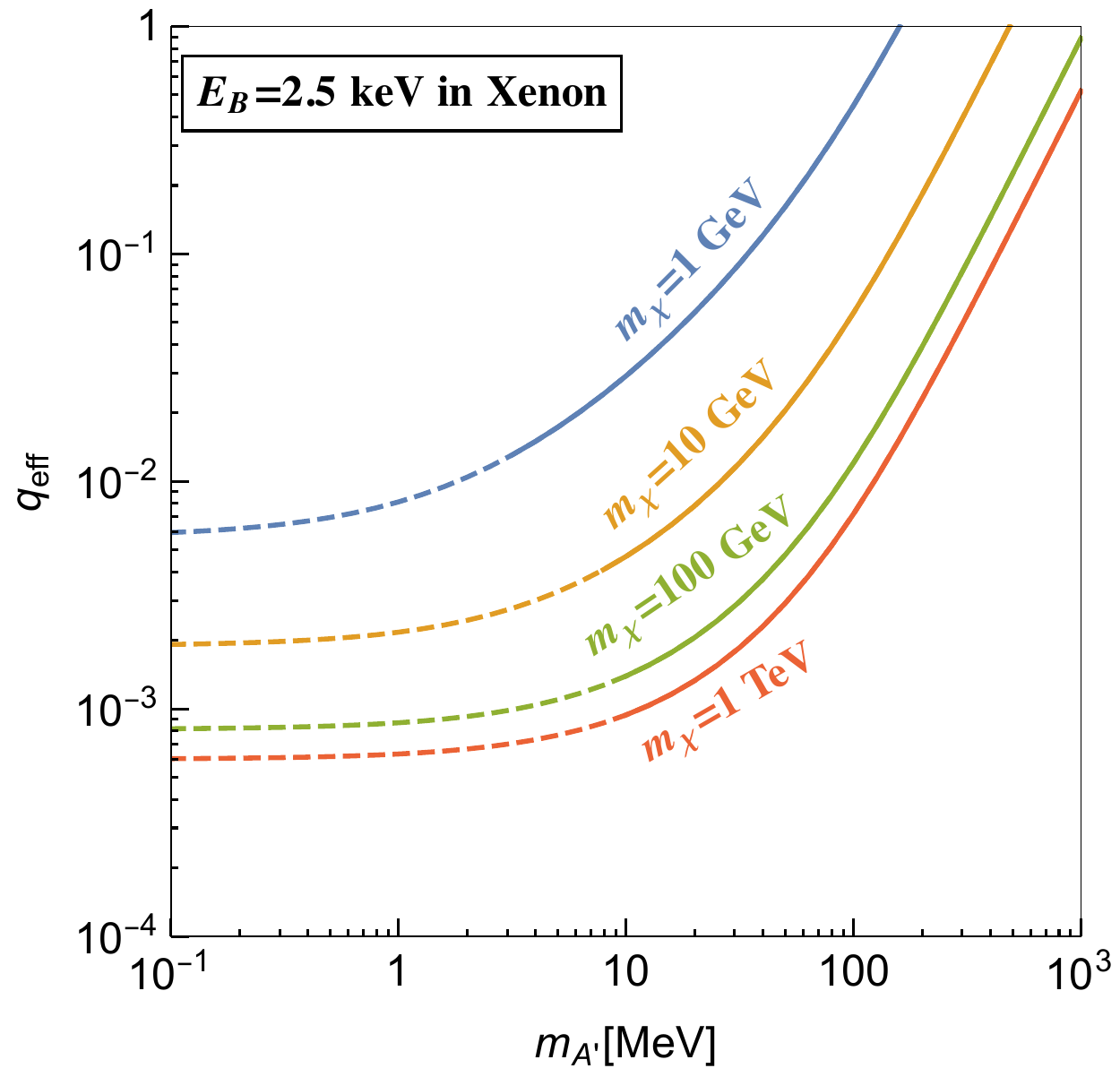}
\caption{In the $\{\mAp, \qeff\}$ plane, contours of fixed binding energy $E_B= \SI{2.5}{\kilo\eV}$ of $(\x\,\text{Xe}^+)$ for different choices of $m_\x$. Along the dashed parts of the contours, $\x$ also binds with Fe.} 
\label{fig:2.5keVXe}
\end{figure}

This leads to the intriguing possibility that binding to heavy elements, such as xenon and thallium, is allowed, while for lighter elements, such as nitrogen and iron, no bound state exists. In this case, $\x$ does not bind to the light elements it encounters when traversing Earth's atmosphere and crust, but does bind to heavy elements employed in direct detection experiments. 
In the Bohr-like regime ($\mAp \ll \qeff Z \alpha \mu$), the binding energy is
\begin{equation}
\label{estimate}
E_B^{(\text{Bohr})} \sim \SI{10}{\kilo\eV} \times  \bigg( \frac{\qeff}{10^{-3}}\bigg)^2 \bigg(\frac{Z}{54}\bigg)^2 \bigg(\frac{\mu}{\SI{122}{\giga\eV}}\bigg)
~,
\end{equation}
thus opening up an opportunity to search for the $\mathcal{O}(\SI{}{\kilo\eV})$ energy release in the formation of such bound states in xenon-based targets. From Eq.~\eqref{Bargmann1}, for $m_\x \gg \SI{100}{\giga\eV}$ this occurs when the dark photon mass is $\mAp \sim \qeff \times \mathcal{O}(10)$ \SI{}{\giga\eV}.  In Fig.~\ref{fig:2.5keVXe}, we show the value of $\qeff$ that is required to achieve a binding energy of $E_B = \SI{2.5}{\kilo\eV}$ with xenon (corresponding to the recoil energy of the observed excess at XENON1T~\cite{Aprile:2020tmw}) as a function of $\mAp$ for various choices of $m_\x$. This is determined by numerically solving the Schr\"{o}dinger equation (see the Supplemental Material for additional details). For $m_\x \lesssim m_\text{Xe}$, larger $\x$ masses enhance $\mu$ and thus require smaller values of $\qeff$ for fixed $E_B$, while this effect saturates for $m_\x \gg m_\text{Xe}$.\footnote{For $m_\x \lesssim \SI{20}{\giga\eV}$ and $\eps \lesssim 10^{-3}$ (see the right panel of Fig.~\ref{fig:fvsq}), a dark sector coupling of $\alpha_D \gtrsim 1$ is necessary for our choice of model parameters, which is reasonable if $\x$ is a composite state.} As expected, the required value of $\qeff$ decreases for longer-ranged dark photons (smaller $\mAp$), saturating once $\mAp$ is smaller than the inverse Bohr radius of the DM-nuclear bound state. However, for even smaller values of $m_{A'}$, $\x$ also binds with iron, the heaviest abundant element in Earth's crust (denoted by dashed lines in Fig.~\ref{fig:2.5keVXe}), thus preventing $\x$ particles from reaching underground detectors. We are thus motivated to consider dark photon masses of $\mAp \sim (10 - 100) \ \text{MeV}$.

\emph{Terrestrial Thermalization.} Before ultimately binding to a heavy nucleus in an underground direct detection experiment, $\x$ thermalizes to terrestrial temperatures after elastically scatters many times off the much lighter elements in Earth's atmosphere and crust. This is governed by the transfer cross section for elastic DM-nuclear scattering $\x N \to \x N$, which in the perturbative Born limit ($\mAp \gg \qeff Z \alpha \mu$) 
is approximately
\begin{align}
\label{sigmael}
\sigma_T^{(\text{Born})} &\simeq \frac{64 \pi \, Z^2 \alpha^2 \, \qeff^2 \, \mu^2}{3 \, \mAp^4}
\nonumber \\
&\simeq \SI{3e-26}{\centi\meter\squared} \times \bigg( \frac{\qeff}{10^{-3}}\bigg)^2 \bigg( \frac{\SI{50}{\mega\eV}}{\mAp}\bigg)^{4}
~,
\end{align}
where in the second line we have taken $m_\x \gg m_N$ and set the nuclear parameters equal to that of silicon, one of the most abundant elements in Earth's crust. Taking a terrestrial silicon density of $n_\text{Si} \sim \SI{e22}{\per\centi\meter\cubed}$, the typical distance $\ell_\text{therm} \sim (m_\x \, m_N / \mu^2) \, (n_\text{Si} \, \sigma_T)^{-1}$ for $\x$ to equilibrate to room temperature is much smaller than $1 \ \text{km}$ for sub-TeV DM masses.

Earth's gravitational field $g$ induces a radially inward bulk flow of the thermalized $\x$ particles. The drift velocity of this flow is parametrically of size $v_\text{drift} \sim (m_\x / \mu) \, g \, \ell_\text{mfp} / v_\text{rel}$, where $\ell_\text{mfp} \sim 1 / (n_N \, \sigma_T)$ is the mean free path for scattering off nuclei $N$ and $v_\text{rel}$ is the relative thermal velocity between $\x$ and $N$. In the parameter space of interest, this drift is very slow, e.g., $v_\text{drift} \ll 10^{-10}$ for sub-TeV DM masses. Since the virialized galactic $\x$ population continually bombards Earth with a characteristic speed set by the much faster galactic wind $v_\text{vir} \sim 10^{-3}$, conservation of DM flux implies that the terrestrial energy density $\rho_\x$ is greatly enhanced\footnote{The rate to form bound $(\x \bar{\x})$ states on Earth is much too slow to affect the signals discussed here. This is because although such dark interactions are unsuppressed by $\eps \ll 1$, the largest terrestrial $\x$ densities that we consider in this work are  smaller than the density of normal matter by at least thirteen orders of magnitude.} compared to the galactic density~\cite{Pospelov:2020ktu}. Specifically, $\rho_\x \sim (v_\text{vir} / v_\text{drift}) \, f_\x \, \rho_{_\text{DM}}$, where $\rho_{_\text{DM}} \simeq \SI{0.3}{\giga\eV\per\centi\meter\cubed}$ is the local DM energy density.\footnote{For our choice of model parameters, if $m_\x \gg 10 \ \text{TeV}$, then the thermalization distance $\ell_\text{therm}$ is much larger than the detector depth $h$ (which consists of $h \simeq 1.4 \ \text{km}$ of rock at Gran Sasso). As a result, $\x$ particles bombarding the Earth from above do not thermalize before reaching the detector. However, if $\ell_\text{therm}$ is much smaller than the radius of the Earth, $\x$ particles approaching the Earth from below travel more distance through Earth's crust and hence can thermalize above the detector. In this case, the local value of $\rho_\x$ is reduced by the small region of solid angles corresponding to such trajectories, suppressing the thermalized $\x$ density at an underground detector by $\sim h / (2 \ell_\text{therm})$.} Taking the limestone rock of the Gran Sasso overburden to be composed of an equal mixture of calcium carbonate and magnesium carbonate with density $\sim \SI{3}{\gram\per\centi\meter\cubed}$~\cite{Miramonti:2005xq}, we find that the terrestrial number density of $\x$ is approximately
\begin{equation}
n_\x \sim f_\x \times 10^8 \ \text{cm}^{-3} \times
\begin{cases}
\big(\frac{\text{GeV}}{m_\x}\big)^{1.5} & (m_\x \ll m_N)
\\
\big(\frac{\text{GeV}}{m_\x}\big)^{2.1} & (m_\x \gg m_N)
~.
\end{cases}
\end{equation}
In performing this calculation, instead of the Born-like estimate in Eq.~(\ref{sigmael}), we have incorporated a full thermal average of the scattering rate utilizing the semi-analytic results outlined in Ref.~\cite{Colquhoun:2020adl}. A more detailed discussion will be provided in forthcoming work~\cite{future}. 

\emph{Rate of Bound State Formation.} The formation of DM-nuclear bound states occurs by the Migdal/Auger-like ejection of an atomic electron, $\x + A \to (\x A^+) + e^-$, followed by subsequent relaxation of the non-ejected atomic electrons to the ground state. Bound state formation by bremsstrahlung emission of a final state photon, $\x + A \to (\x A) + \gamma$, is subdominant, as is the case for the elastic scattering of light DM, since the rate is suppressed by $\sim \SI{}{\kilo\eV} / m_N$~\cite{Bell:2019egg, Ibe:2017yqa}.

We estimate the probability of the transition $\x + A \to (\x A^+) + e^-$ using quantum mechanical perturbation theory. In the Supplemental Material, we derive the cross section for this process. We focus on the $s$-wave DM-nuclear  final state since it is guaranteed to exist if a bound state is allowed. For an incoming DM particle that is also $s$-wave, the cross section for this $s$-wave to $s$-wave transition is  
\begin{equation}
\sigma_B v_\text{rel} \simeq \frac{4 \pi}{9} \, \frac{(Z \alpha \, m_e)^5}{(2 \mu \, E_B)^{7/2}} \, \Big(\frac{\mu}{m_N} \Big)^4 \, F_\x^2 \, F_e^2
~.
\end{equation} 
The numerical factors $F_\x$ and $F_e$ depend on the initial and final state wavefunctions of $\x$ and the ejected electron, respectively, and must be determined numerically. For a binding energy of $E_B = \SI{2.5}{\kilo\eV}$ in xenon, we find $F_e \simeq 0.7$. Additionally fixing $\qeff$ as in Fig.~\ref{fig:2.5keVXe} and $\mAp = \SI{50}{\mega\eV}$, we find that $F_\x \sim \mathcal{O}(10)$ with the precise value depending on the DM mass (see the Supplemental Material for additional details). Upon evaluating $F_\x$ and $F_e$, we find that the cross section for forming DM-xenon bound states with $E_B = 2.5 \ \text{keV}$ and $\mAp = \SI{50}{\mega\eV}$ is well fit by the functional form
\begin{equation}
\sigma_B v_\text{rel} \simeq \SI{6e-34}{\centi\meter\squared} \times \left(\frac{\mu}{100 \ \text{GeV}}\right)^{0.55}
~.
\end{equation}
In the Supplemental Material, we show that although $p$-wave (and higher) to $s$-wave transitions also occur, they are suppressed by the small temperature of the thermalized DM population and are thus subdominant to the $s$-wave to $s$-wave transition discussed above. 

\emph{Recombination Signal.} Direct detection experiments that employ heavy elements, such as liquid noble targets, have remarkable sensitivity to the formation of these DM-nuclear bound states. For a detector employing a target of atomic mass $A$, the signal rate per unit target mass per unit time  is
\begin{equation}
R_\text{sig} = P_\text{surv} \, n_\x \, \sigma_B v_\text{rel} \, \left(\frac{N_A}{A\ \text{grams}} \right)
~,
\end{equation}
where $N_A$ is Avogadro's number and $P_\text{surv}$ is the survival probability that an incoming $\x$ is not captured by naturally occurring elements in the terrestrial overburden. Since we are interested in dark photon masses that forbid $\x$ from binding with elements much lighter than xenon, premature capture in the terrestrial crust can only occur from binding with rare heavy elements. 

The most abundant of such elements is barium, which possesses an atomic number and mass slightly greater than xenon. The average density of barium in Earth's crust is $n_\text{Ba} \sim \SI{e18}{\per\centi\meter\cubed}$ with an exact value that varies geographically. For our estimates below, we adopt a range $n_\text{Ba} \in (0.7 - 6) \times \SI{e18}{\per\centi\meter\cubed}$, whose log-central value of $2 \times \SI{e18}{\per\centi\meter\cubed}$ is representative of limestone~\cite{wedepohl2014elements}, the dominant rock in the overburden at Gran Sasso National Laboratory. The survival probability is then given by $P_\text{surv} \simeq \text{exp}[- n_\text{Ba} \, \sigma_\text{cap} v_\text{rel} \, \Delta t]$, where $\sigma_\text{cap}$ is the cross section for $\x$ to bind with barium and $\Delta t$ is the time it takes for $\x$ to penetrate the XENON1T overburden, which consists of $h \simeq 1.4 \ \text{km}$ of rock. The transit time is given by the minimum of either the diffusion or gravitational-drift timescales, $\Delta t \sim \min (t_\text{diff} , t_\text{drift})$, where the diffusion time $t_\text{diff}$ and the gravitational-drift timescale $t_\text{drift}$ are approximately
\begin{equation}
t_\text{diff} \sim h^2 \, \frac{\Gamma_\text{drag} \, m_\x}{\SI{300}{\kelvin}}
~~,~~
t_\text{drift} \sim h \, \frac{\Gamma_\text{drag}}{g}
~,
\end{equation}
and $\Gamma_\text{drag} \sim (\mu / m_\x) n_N \sigma_T v_\text{rel}$ is the drag rate (the inverse timescale for a particle to change its momentum by an $\mathcal{O}(1)$ amount) from elastic scattering off of light nuclei with density $n_N$ in the crust. Fixing $E_B = 2.5 \ \text{keV}$ in xenon and $\mAp = 50 \ \text{MeV}$, we find that diffusion is efficient ($t_\text{diff} \ll t_\text{drift}$) for $m_\x \ll \SI{100}{\giga\eV}$, while gravitational drift is most efficient ($t_\text{drift} \ll t_\text{diff}$) for $m_\x \gg \SI{100}{\giga\eV}$, such that $P_\text{surv} \simeq 1$ in either case. Instead, for DM masses of $m_\x \sim \SI{100}{\giga\eV}$, $P_\text{surv} \ll 1$ and premature capture by naturally occurring barium may exponentially reduce the DM flux that is able to reach underground detectors.\footnote{Even if capture by barium is significant, a flux of DM particles can be inadvertently delivered to a lab through other means, such as air ventilation, bypassing the need to diffuse through the rock. We conservatively neglect such processes in our analysis.}

\begin{figure*}
\hspace*{-0.2in}
\vspace*{0.5cm}
\centering
  \includegraphics[width=0.47\textwidth]{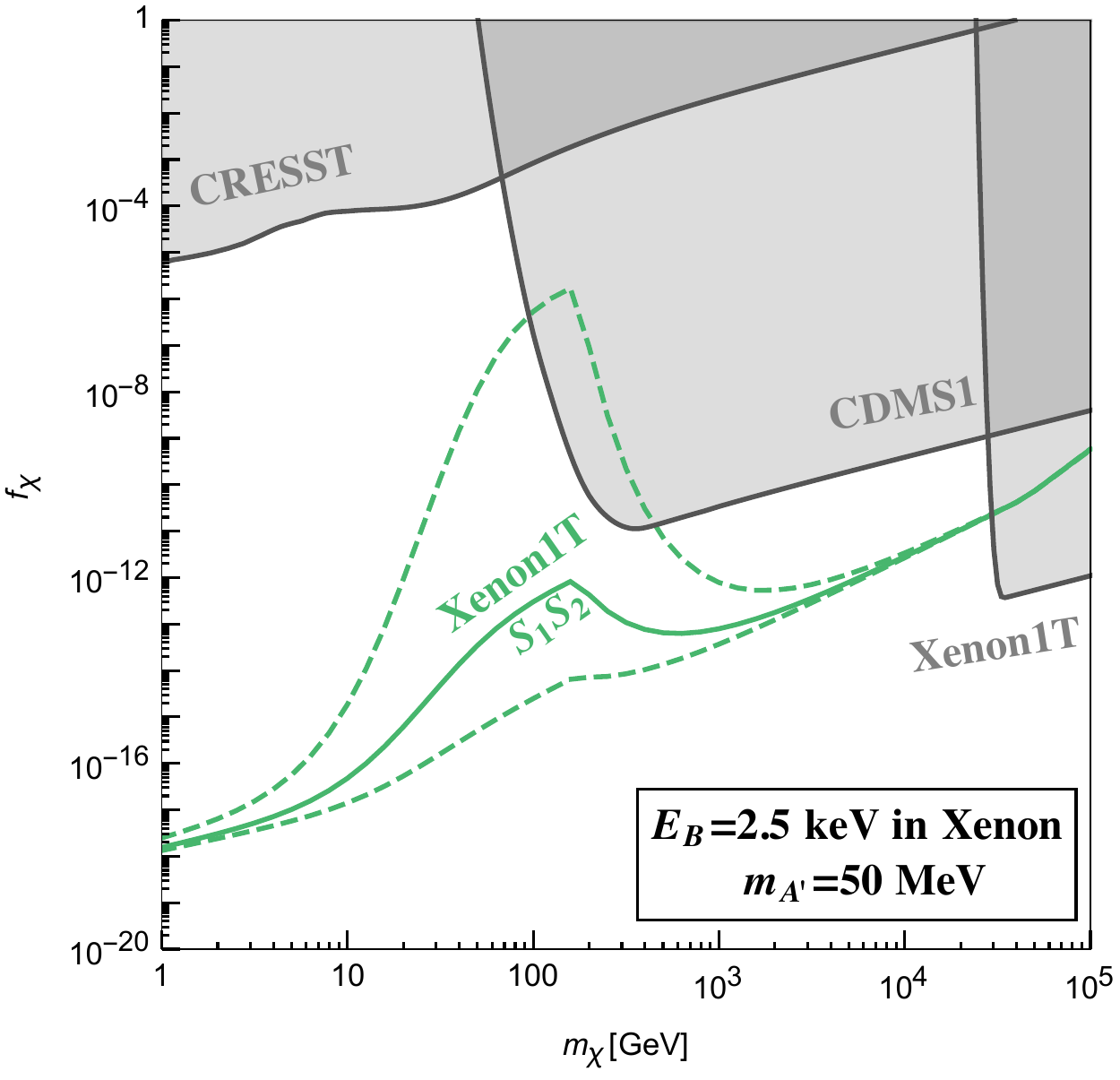}
  \includegraphics[width=0.47\textwidth]{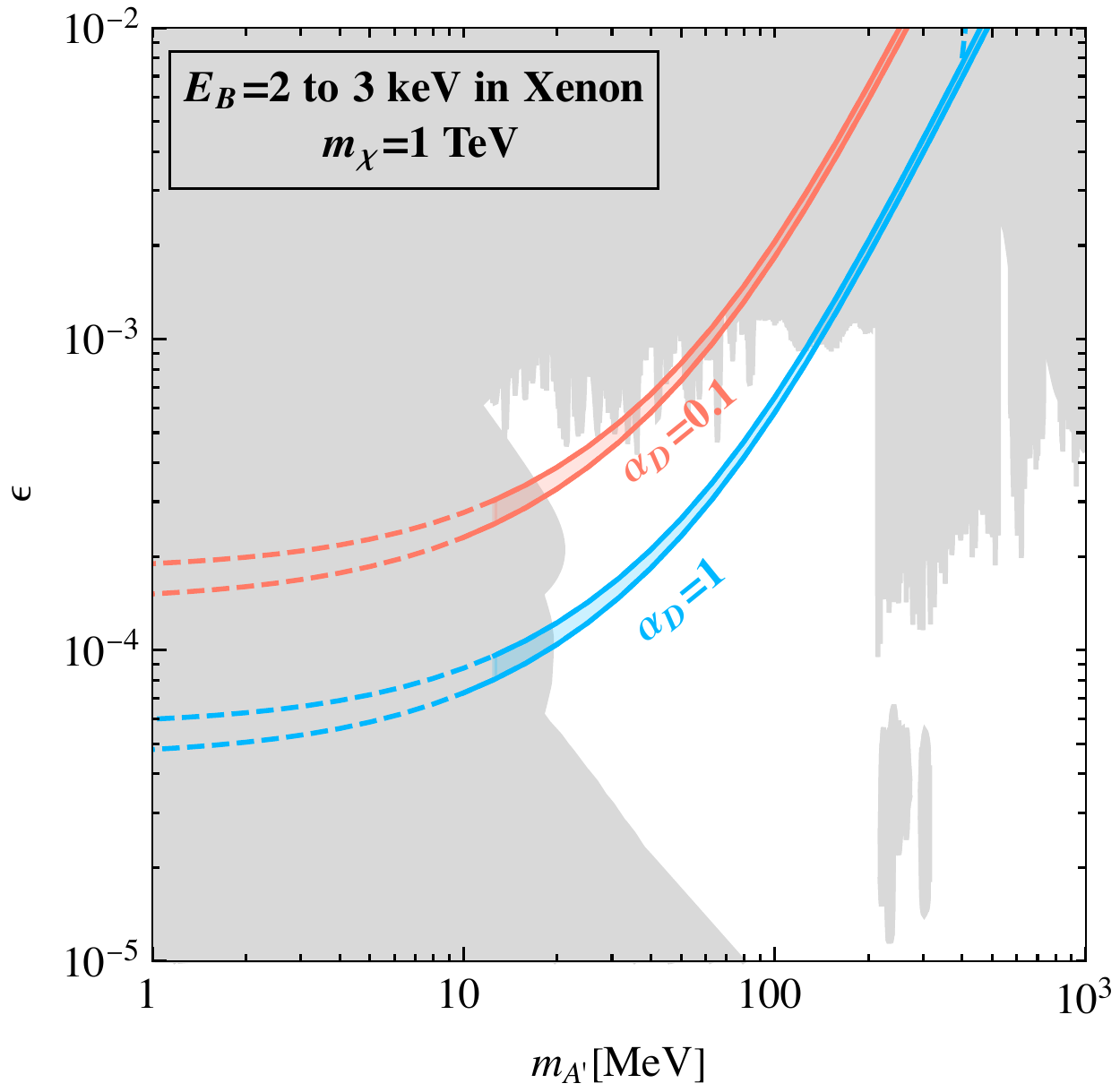}
  \caption{\textit{(Left)} Limits derived from XENON1T (green) on the fractional abundance $f_\x$, in the case that the binding energy between $\x$ and xenon is $E_B = \SI{2.5}{\kilo\eV}$ and $\mAp = \SI{50}{\mega\eV}$. The solid green line assumes an abundance of barium in the Gran Sasso overburden of $2 \times \SI{e18}{\per\centi\meter\cubed}$, whereas the upper and lower dashed green lines assume a barium density that is larger or smaller by a factor of three, respectively. Also shown are existing constraints (shaded gray) from searches for $\x$-nuclear elastic scattering at CRESST~\cite{CRESST:2017ues}, CDMS~\cite{CDMS:2002moo}, and XENON1T~\cite{XENON:2017vdw}. \textit{(Right)} Contours of $(\x\,\text{Xe}^+)$ binding energy $\SI{2}{\kilo\eV} \le E_B \le \SI{3}{\kilo\eV}$ for various choices of $\alpha_D$, fixing $m_\x = 1 \ \text{TeV}$, in the $\{\mAp, \epsilon\}$ plane. Along the dashed parts of the contours, $\x$ also binds with Fe. The gray regions are currently excluded by accelerator and beam dump searches for visibly decaying dark photons~\cite{Lanfranchi:2020crw}. Future acclerator searches, for instance   APEX~\cite{APEX:2011dww}, FASER~\cite{FASER:2018eoc}, HPS~\cite{HPS:2018xkw}, LHCb~\cite{Ilten:2015hya,Ilten:2016tkc}, NA64~\cite{Gninenko:2013rka,Andreas:2013lya}, and SeaQuest~\cite{Berlin:2018pwi}, will explore nearly all of the currently allowed parameter space shown.}
   \label{fig:fvsq}
\end{figure*}

The time dependence of these signals depends on the terrestrial $\x$ density, which scales linearly with the velocity of the virialized galactic population as $n_\x \propto v_\text{vir}$. Since in Earth's frame $v_\text{vir}$ varies due to the relative motion of Earth around the Sun, we expect an annual modulation of $n_\x$. Furthermore, the short mean free path in Earth's crust implies a sizeable diurnal modulation. For the signals discussed in this work, modulation over a timescale $t_\text{mod}$ is not present if the diffusion time for $\x$ to travel a length $v_\text{drift} \, t_\text{mod}$ is shorter than $t_\text{mod}$. We estimate that for masses $m_\x \ll 10 \ \text{GeV}$ or $m_\x \ll 100 \ \text{GeV}$, diffusion is strong enough to wash out an annual or daily modulation in XENON1T data, respectively. Since, unlike a WIMP, $\x$ does not free stream throughout Earth, the phase of the annual modulation depends on the relative direction between the vector normal to Earth's surface at Gran Sasso and Earth's galactic motion. As a result, we find that the annual modulation for such strongly interacting relics peaks near the spring and fall equinox, which is consistent with the findings of Ref.~\cite{Laletin:2019qca}. 

In Fig.~\ref{fig:fvsq} (left), we illustrate the existing sensitivity of XENON1T to fractional abundances $f_\x$ as a function of $m_\x$, fixing $\mAp = 50 \ \text{MeV}$ and $E_B = \SI{2.5}{\kilo\eV}$ in xenon. Following Ref.~\cite{Bloch:2020uzh}, an S1-S2 signal rate of $R_\text{sig} \simeq 60 \ \text{tonne}^{-1} \ \text{yr}^{-1}$ accounts for the excess of electron recoil events as measured by XENON1T. This model space is shown as the green lines in Fig.~\ref{fig:fvsq} (left) for various assumptions regarding the possible barium abundance in the Gran Sasso overburden (from bottom to top, these contours assume a barium density of  $0.7 \times \SI{e18}{\per\centi\meter\cubed}$, $2 \times \SI{e18}{\per\centi\meter\cubed}$, and $6 \times \SI{e18}{\per\centi\meter\cubed}$, respectively). Thus, values of $f_\x$ that lie significantly above these lines are presently excluded by XENON1T. We see that if a component of DM can bind to heavy nuclei, XENON1T probes fractional abundances as small as $f_\x \sim 10^{-18}$. Previously existing constraints on this parameter space are also shown as shaded gray regions. These are derived from searches at a surface-level CRESST run~\cite{CRESST:2017ues}, as well as underground CDMS~\cite{CDMS:2002moo} and XENON1T~\cite{XENON:2017vdw} runs that are sensitive to the nuclear elastic scattering of $\x$ particles that do not thermalize with the terrestrial environment. For small $\x$ masses, these limits are significantly weakened due to the increased likelihood for $\x$ to thermalize in the overburden, and thus are typically many orders of magnitude weaker than those derived from bound state formation in XENON1T. 

We also note that a search for anomalously heavy gold atoms (in this case, due to heavy $\x$ particles bound to the nucleus) can be recast as constraints in this model space~\cite{Javorsek:2002bf}. Conservatively assuming that the gold nuggets used in these tests originated below the point of thermalization for $\x$ throughout the entire age of the Earth, and that the capture rate in gold is similar to that of xenon, the corresponding limits are subdominant to the direct detection constraints described above. In particular, we estimate that these searches are sensitive to fractional abundances of $f_\x \gtrsim 10^{-9}$ for $m_\x \sim \SI{100}{\giga\eV}$ and $f_\x \gtrsim 10^{-6}$ for $m_\x \sim \SI{}{\tera\eV}$, with no sensitivity outside this mass range.  

\emph{Discussion.} We have explored the intriguing possibility that the excess of events at XENON1T is explained by an exponentially small subcomponent of DM that binds with heavy nuclei through a dark photon mediator of mass $\mAp \sim \mathcal{O}(10-100) \ \text{MeV}$ that generates a sizeable DM-SM coupling $\qeff = \eps \sqrt{\alpha_D / \alpha} \gtrsim 10^{-3}$. For a perturbative model in which $\alpha_D \lesssim \mathcal{O} (1)$, we are thus motivated to consider $\eps \gtrsim 10^{-4}$. This parameter space can be efficiently probed by searching for visible decays of dark photons that are produced in the collisions of dedicated accelerator experiments.\footnote{For $\mAp \ll 1 \ \text{MeV}$, accelerator searches for GeV-scale $\x$ particles exclude $\qeff \gtrsim 10^{-2}$~\cite{ArgoNeuT:2019ckq}, but lack sensitivity to models with much larger $\Ap$ masses, as considered in this work.} This is illustrated in Fig.~\ref{fig:fvsq} (right) in the $\{\mAp,\varepsilon\}$ plane. For two representative choices of $\alpha_D$ shown as red and blue bands, the $\x-\text{Xe}$ binding energy is varied from $(2 - 3) \ \text{keV}$, while fixing $m_\x = \SI{1}{\tera\eV}$. Also shown in dark gray are existing constraints from searches for the production and visible decay of dark photons in accelerator and beam dump experiments~\cite{Lanfranchi:2020crw}. Near future accelerator experiments, such as APEX~\cite{APEX:2011dww}, FASER~\cite{FASER:2018eoc}, HPS~\cite{HPS:2018xkw}, LHCb~\cite{Ilten:2015hya,Ilten:2016tkc}, NA64~\cite{Gninenko:2013rka,Andreas:2013lya}, and SeaQuest~\cite{Berlin:2018pwi}, will be able to decisively probe this region of parameter space~\cite{Agrawal:2021dbo}. 

We have focused on exhibiting the salient features of how DM-SM bound state formation can give rise to signals in the XENON1T  S1-S2 data of Ref.~\cite{Aprile:2020tmw} since this search has the largest exposure among experiments sensitive to $\mathcal{O} (\text{keV})$ electron recoils. However, other experiments may have sensitivity to new parameter space for two reasons. First, when threshold energies are reduced to below $\sim 1 \ \text{keV}$ at the expense of reduced exposure or larger backgrounds, smaller couplings $\qeff$ can be probed for larger $f_\x$. Examples in this vein are the XENON1T S2-only~\cite{XENON:2019gfn} and XENON10~\cite{XENON10:2011prx} searches, which are sensitive to energy depositions as small as $\mathcal{O}(100) \ \text{eV}$ and $\mathcal{O}(10) \ \text{eV}$, respectively. Second, nuclear targets which are heavier than xenon are sensitive to smaller $\qeff$. Examples among these are tungsten in the CRESST experiment~\cite{CRESST:2017ues} and thallium dopant present in the DAMA experiment~\cite{DAMA:2008jlt,DAMA:2010gpn,Bernabei:2018jrt}. However, since CRESST vetoes on electron recoil events and reports limits only on nuclear recoil events, its public results are at present insensitive to this model. 

The DAMA experiment has reported a long-standing annual modulation signal in the $(1-6) \ \text{keV}$ energy range~\cite{DAMA:2008jlt,DAMA:2010gpn,Bernabei:2018jrt}. While the primary target materials (sodium and iodine) are lighter than xenon, thallium is present at the $\sim 10^{-3}$ level. Since thallium nuclei are much heavier than xenon, this raises the tantalizing possibility that $\qeff$ is large enough such that $\x$ binds deeply in thallium but is too small to deposit a significant energy above threshold in xenon-based experiments, thus explaining the DAMA observation. For $m_\x \gtrsim 100 \ \text{GeV}$, explaining the modulated rate at DAMA requires fractional abundances of $f_\x \gtrsim 10^{-9}$, which is in conflict with  limits derived from CDMS~\cite{CDMS:2002moo} (see Fig.~\ref{fig:fvsq} (left)). Instead, for $m_\x \lesssim 100 \ \text{GeV}$, $f_\x \sim 10^{-12} - 10^{-9}$ predicts a rate consistent with DAMA, provided that $\qeff$ lies in a narrow range to facilitate a $\x-\text{Tl}$ binding energy of $E_B\sim 2 \ \text{keV}$ without being in tension with an S2-only search at XENON1T~\cite{XENON:2019gfn}. However, as discussed above, the phase of the annual modulation of these signals is inconsistent with DAMA's observation of peak rates near the beginning of June and December. Although strongly interacting relics cannot accommodate this excess of events, it is interesting to note that DAMA can place some of the most stringent constraints on models in which $\x$ preferentially binds to thallium (and heavier elements).

\emph{Acknowledgements.} The authors would like to acknowledge Carlos Blanco, Oren Slone, and Neal Weiner for fruitful conversations. AB is supported
by the James Arthur Fellowship. HL is supported by the DOE under Award Number DE-SC0007968, NSF grant PHY-1915409, and the Simons Foundation.   MP is supported in part by U.S. Department of Energy (Grant No. DE-SC0011842). HR is supported in part by NSF Grant PHY-1720397 and the Gordon and Betty Moore Foundation Grant GBMF7946.

\bibliographystyle{apsrev4-1}
\bibliography{biblio}

\clearpage
\newpage
\maketitle
\onecolumngrid
\begin{center}
\textbf{\large Low-Energy Signals from the Formation of Dark Matter-Nuclear Bound States} \\ 
\vspace{0.05in}
{ \it \large Supplemental Material}\\ 
\vspace{0.05in}
{}
{Asher Berlin, Hongwan Liu, Maxim Pospelov, and Harikrishnan Ramani}

\end{center}
\setcounter{equation}{0}
\setcounter{figure}{0}
\setcounter{table}{0}
\setcounter{section}{1}
\renewcommand{\theequation}{S\arabic{equation}}
\renewcommand{\thefigure}{S\arabic{figure}}
\renewcommand{\thetable}{S\arabic{table}}
\interfootnotelinepenalty=10000 


In this Supplemental Material, we provide a detailed derivation of the cross section for bound state formation $\x + A \to (\x A^+) + e^-$.

\section*{Perturbation Theory Set-Up}

The Hamiltonian for the system can be written as $\hat{H} = \hat{H}_0 + \hat{V}_\text{pert}$, with $\hat{V}_\text{pert}$ being a small perturbation to the unperturbed Hamiltonian $\hat{H}_0$. The cross section for transitioning from an initial state $\ket{i}$ to a continuum final state $\ket{f}$, both eigenstates of $\hat{H}_0$, is given by Fermi's golden rule. In the process we are considering, an electron is ejected into a continuum final state, with a change in energy given by $E_B$, the binding energy of $(\x A)$, so that Fermi's golden rule reads~\cite{Landau1981Quantum}
\begin{alignat}{1}
    \dd \sigma =  \frac{\dd ^3 \mathbf{p}_e}{(2\pi)^3} \, 2 \pi \, \left| \braket{f | \hat{V}_\text{pert} | i} \right|^2 \delta \left(E_e - E_B + \omega_{e,i}  \right) = \frac{\dd \Omega_e }{(2\pi)^2} \, m_e p_e \, \left| \braket{ f | \hat{V}_\text{pert} | i} \right|^2  \,,
    \label{eq:fermis_golden_rule}
\end{alignat}
where $\mathbf{p}_e$ and $E_e$ are the momentum and energy of the final state electron respectively, and $\omega_{e,i}$ is the binding energy of the initial state electron. In the second equality, we have integrated over the delta function, which leaves just the solid angle of the electron momentum $\Omega_e$ and fixes $p_e = \sqrt{2 m_e (E_B - \omega_{e,i})}$. We neglect the kinetic energy of the incoming thermalized $\x$ particle, which is much smaller than $E_B$. The appropriate normalization of the states $\ket{i}$ and $\ket{f}$ will be discussed below. 

The total Hamiltonian involving the nucleus $N$, the DM particle $\x$, and the ejected electron $e$ contains the following terms:
\begin{alignat}{1}
    \hat{H} = \hat{T} + \hat{V}_{\x N} 
    + \hat{V}_{e A} \,, 
\end{alignat}
where $\hat{T}$ is the kinetic energy terms of all the particles involved, and the various potential energy contributions are $\hat{V}_{\x N}$ between $\x$ and the nucleus $N$ and $\hat{V}_{eA}$ between the ejected electron and the atom (including both $N$ and the other unejected electrons).\footnote{The other potential energy terms involving solely the unejected electrons are unimportant, since either the ejected electron states or the $\x$ states are eigenstates of these operators, and the initial and final wavefunctions are orthogonal. Furthermore, we do not include the potential arising from interactions between $\x$ and electrons because $\x$ becomes tightly bound to the atom and thus dominantly experiences just the bare nucleus: the typical size of the bound state, given in Eq.~\eqref{eq:bound_state_size}, is much smaller than the size of the xenon atom.} Let us now discuss both of these contributions.

We model $V_{eA}$ by adopting the Thomas-Fermi model for the neutral atom, which gives an effective screened potential energy of the nucleus and the atomic electrons as a function of the distance to the nucleus. In this model, the effective screened potential energy of the entire atom $V_A$ is~\cite{Landau1981Quantum}
\begin{alignat}{1}
    V_A(|\mathbf{r}_e - \mathbf{r}_N|) = - \, \frac{Z \alpha}{|\mathbf{r}_e - \mathbf{r}_N|} ~ \phi \left(\frac{|\mathbf{r}_e - \mathbf{r}_N|}{b}\right) \,,
\end{alignat}
where $b \equiv (9 \pi^2 / 2Z)^{1/3} / (4 \alpha m_e)$, and $\phi(\xi)$ satisfies the Thomas-Fermi equation,
\begin{alignat}{1}
    \frac{d^2 \phi}{d \xi^2} = \frac{\phi^{3/2}}{\sqrt{\xi}} \,,
    \label{eq:thomas_fermi_eqn}
\end{alignat}
with the boundary conditions $\phi(0) = 1$ and $\lim_{\xi \to \infty} \phi(\xi) = 0$, which can be solved numerically. $V_{e A}$, the potential energy between the ejected electron and the rest of the atom, can be obtained by subtracting the contribution from the initial bound state electron, i.e., 
\begin{alignat}{1}
    V_{e A}(|\mathbf{r}_e - \mathbf{r}_N|) = V_A(|\mathbf{r}_e - \mathbf{r}_N|) - \alpha \int \dd ^3 \mathbf{r}' \frac{|\psi_{e,i}(\mathbf{r}')|^2}{|(\mathbf{r}_e - \mathbf{r}_N) - \mathbf{r}'|} \,,
\end{alignat}
where $\psi_{e,i}(\mathbf{r})$ is the wavefunction of the initial bound state electron before it is ejected. The explicit form of $\psi_{e,i}(r)$ will be discussed below near Eq.~(\ref{eq:Psiei}). 

The characteristic size of the $(\x N)$ bound state is
\begin{alignat}{1}
    \frac{1}{\sqrt{2 \mu E_B}} \simeq \SI{9}{\femto\meter} \times \left(\frac{\SI{100}{\giga\eV}}{\mu}\right)^{1/2} \left(\frac{\SI{2.5}{\kilo\eV}}{E_B}\right)^{1/2} \,, 
    \label{eq:bound_state_size}
\end{alignat}
which is comparable to typical nuclear radii $R_\text{nuc} \sim \SI{5}{\femto\meter} \times (A/100)^{1/3}$. Here, $\mu$ is the reduced mass of $\chi$ and $N$. Finite size effects of the nucleus are therefore important for describing the potential energy $V_{\x N}$ between $\x$ and $N$. Modeling the nucleus as a uniformly charged sphere of radius $R_\text{nuc}$ centered at $\mathbf{r}_N$, the potential at $\mathbf{r}_\x$ can be written as
\begin{alignat}{1}
    V_{\x N} (|\mathbf{r}_\x - \mathbf{r}_N|) = - \int_N \dd^3 \mathbf{r}' ~ \frac{\rho}{|(\mathbf{r}_\x - \mathbf{r}_N) - \mathbf{r}'|} ~ e^{-\mAp|(\mathbf{r}_\x - \mathbf{r}_N) - \mathbf{r}'|}
    ~,
\end{alignat}
where $\rho \equiv q_\text{eff} \, Z \, \alpha \, / (4 \pi R_\text{nuc}^3 / 3)$, and the integral is performed over the volume of the nucleus. This integral can be evaluated analytically to give
\begin{alignat}{1}
    V_{\x N}(r) = - \, \frac{4 \pi \, \rho}{\mAp^3 \, r}  \begin{cases}
        \mAp \, r - e^{- \mAp R_\text{nuc}} (1 + \mAp \, R_\text{nuc}) \, \sinh \left(\mAp \,  r\right) & (r < R_\text{nuc}) \\
        e^{-\mAp r} \Big[\mAp \, R_\text{nuc} \cosh (\mAp \, R_\text{nuc}) - \sinh(\mAp \, R_\text{nuc}) \Big] & (r \geq R_\text{nuc}) \,,
    \end{cases}
\end{alignat}
where we have defined $\mathbf{r} \equiv \mathbf{r}_N - \mathbf{r}_\x$. Defining the origin to be located at the $(\chi N)$ center-of-mass, $\mathbf{r}_N = (\mu / m_N) \, \mathbf{r}$ and $\mathbf{r}_\x = -(\mu / m_\x) \,  \mathbf{r}$. Hence, after bound state formation, the position of $N$ is parametrically $\mu/m_N$ times the spatial extent of the bound state, i.e., $r_N \sim (\mu / m_N) / \sqrt{2 \mu E_B}$. Since this is always much smaller than the typical size of the atom, we may expand $V_{e A}(|\mathbf{r}_e - \mathbf{r}_N|)$ in powers of the small quantity $\mathbf{r}_N$ to obtain
\begin{align}
    V_{e A}(|\mathbf{r}_e - \mathbf{r}_N|) 
    &= V_{e A}(r_e) + \mathbf{r}_N^i \, \partial^i \, V_{e A}(r_e) + \frac{1}{2} \, \mathbf{r}_N^i \mathbf{r}_N^j \, \partial^i \partial^j \, V_{e A}(r_e) + \cdots
    \nonumber \\
    &\equiv V_{e A}^{(0)} + V_{e A}^{(1)} + V_{eA}^{(2)} + \cdots 
    ~,
\end{align}
where $i$, $j$ denote spatial components, with repeated indices summed over. We now take the unperturbed Hamiltonian to be
\begin{alignat}{1}
    \hat{H}_0 \equiv \hat{T} + \hat{V}_{\x N} + \hat{V}_{e A}^{(0)} \,. 
\end{alignat}
The initial and final states of both $\x$ and the ejected electron $e$ are bound states or asymptotically free states that are eigenstates of $\hat{H}_0$; the actual wavefunctions will be worked out below. The full perturbative portion of the Hamiltonian is therefore $\hat{V}_{e A}^{(1)} + \hat{V}_{e A}^{(2)} + \cdots $. We therefore define the perturbative correction to the Hamiltonian as
\begin{alignat}{1}
    \hat{V}_\text{pert} \equiv \hat{V}_{eA}^{(1)} + \hat{V}_{eA}^{(2)} \,,
\end{alignat}
keeping just the first two terms in the expansion of $V_{e A}(|\mathbf{r}_e - \mathbf{r}_N|)$. We assume that $\x$ is captured into the $s$-wave ground state of the potential $V_{\x N}$, since it is the most deeply bound; as we show below, under this assumption, $\hat{V}_{eA}^{(1)}$ leads to a $p$-to-$s$ transition with a unit change in angular momentum between the initial and final $\x$ states, while $\hat{V}_{eA}^{(2)}$ facilitates an $s$-to-$s$ transition instead, with no change in $\x$ angular momentum. We compute the rate for each angular state by incorporating each contribution to $\hat{V}_\text{pert}$ separately, estimating the total bound state capture cross section as $\sigma_B \approx \sigma_s + \sigma_p$, where $\sigma_{s,p}$ is the cross section for an $s$-to-$s$ or $p$-to-$s$ transition, respectively. This treatment is appropriate as long as $\sigma_s \gg \sigma_p$ or vice-versa, which is true across almost all relevant parameter space.

\section*{Wavefunctions in a Central Potential}

Before we investigate each transition separately, we note that $\x$ and $e$ are both subject to a two-body central potential with the nucleus. In each case, the Schr\"{o}dinger equation is separable into an angular and a radial component. For a central potential $V$, and a well-defined energy $E$ and angular momentum quantum number $l$, the radial wavefunction $R (\rsc)$ satisfies the following equation: 
\begin{alignat}{1}
    \frac{1}{\rsc^{\, 2}} \, \frac{\dd}{\dd \rsc} \left(\rsc^{\, 2} \, \frac{\dd R}{\dd \rsc}\right) - \left[ \frac{l(l+1)}{\rsc^{\, 2}} + 2 \mu \, V \right] R = -2 \mu E \, R
    ~,
    \label{eq:schrodinger_radial}
\end{alignat}
where $\rsc$ is the relative coordinate between the two bodies, and $\mu$ is the reduced mass of two bodies. For a bound state wavefunction $E < 0$, whereas for a continuum wavefunction $E \simeq k^2 / 2 \mu > 0$ such that $\mathbf{k}$ is the momentum of the particle far away from the nuclear potential.

For bound wavefunctions, the full wavefunction is $R$ times the appropriate spherical harmonic for angular momentum $l$ and azimuthal quantum number $m$. Instead, continuum wavefunctions with definite total momentum $\mathbf{k}$ can be expanded as~\cite{Landau1981Quantum}
\begin{alignat}{1}
    \psi (\rsc \, , \theta) = \sum_{l=0}^\infty \,  \frac{2l + 1}{2k} \, i^l \, P_l(\cos \theta) \, R_{kl}(\rsc) \,,
    \label{eq:continuum_psi_basis}
\end{alignat}
where $R_{kl}(\rsc)$ is the solution to Eq.~\eqref{eq:schrodinger_radial} with 
$E = k^2 / 2 \mu$, $P_l$ is the $l^\text{th}$ Legendre polynomial, and $\cos{\theta} \equiv \hat{\rsc \, } \cdot \hat{\mathbf{k}}$  (we have neglected the phase shift factor $\delta_\ell$, consistent with the perturbative treatment used here). The constant prefactors correctly normalize the asymptotic wavefunction to a single particle plane wave. 

\section*{s-to-s Capture Cross Section}
\label{subapp:s-to-s}

We first consider the case of $s$-to-$s$ capture through $\hat{V}_{e A}^{(2)}$. Although this term is higher order in $r_N$ than $\hat{V}_{e A}^{(1)}$, there is no velocity suppression to this cross section, since there is no change in $l$ between the initial and final states. From Eq.~\eqref{eq:fermis_golden_rule}, the matrix element for this process in the position basis is
\begin{alignat}{1}
\label{eq:dsigmaStoS}
    \dd \sigma = \frac{d\Omega_e }{(2\pi)^2} \, m_e p_e \, \left| \int \dd ^3 \mathbf{r} ~ \psi_{\x,f}^*(\mathbf{r}) \, \psi_{\x,i}(\mathbf{r}) \int \dd ^3 \mathbf{r}_e ~ \psi_{e,f}^*(\mathbf{r}_e) \, \psi_{e,i}(\mathbf{r}_e) ~ V_{e A}^{(2)}(r_e) \right|^2 \,,
\end{alignat}
where the subscripts $i$ and $f$ denote initial and final states for the position-space wavefunctions $\psi$ for $\x$ and $e$. We can simplify this integral with the following identity that we will utilize several times in this Supplemental Material:
\begin{alignat}{1}
    \int \dd ^3 \mathbf{x} ~ f(x) \, \mathbf{x}^i \mathbf{x}^j = \frac{1}{3} \int \dd ^3 \mathbf{x} ~  f(x) \, x^2 \, \delta^{ij} \, ,
    \label{eq:components_in_integral_identity}
\end{alignat}
for some function $f(x)$. Since the $s$-wave initial and final $\x$ states are spherically symmetric, in Eq.~(\ref{eq:dsigmaStoS}) we can therefore replace 
\begin{alignat}{1}
    V_{e A}^{(2)} = \frac{1}{2} \, \mathbf{r}_N^i \mathbf{r}_N^j \, \partial^i \partial^j \, V_{e A}(r_e) \to \frac{r_N^2}{6} \, \nabla^2 V_{e A}(r_e) \,. 
\end{alignat}
By Gauss's law, we can relate $\nabla^2 V_{e A}$ to the charge density of the atom, such that
\begin{alignat}{1}
    V_{e A}^{(2)}(\mathbf{r}_e) \to \frac{2 \pi}{3} \, \alpha \, r_N^2 \big(- Z\delta^3(\mathbf{r}_e) + n_\text{TF}(\mathbf{r}_e) - |\psi_{e,i}(\mathbf{r}_e)|^2 \big) \,,
\end{alignat}
where $n_\text{TF}(\mathbf{r}_e) - |\psi_{e,i}(\mathbf{r}_e)|^2$ is the number density of electrons in the Thomas-Fermi model minus the number density of the initial bound state electron, and the Dirac delta function arises from the nucleus, which we take to be a point charge from the point of view of the ejected electron. The expression for the electronic number density $n_\text{TF}$ is~\cite{Landau1981Quantum}
\begin{alignat}{1}
    n_\text{TF}(\mathbf{r}_e) = \frac{1}{3 \pi^2} \left[\frac{2 \, Z \alpha \, m_e}{r_e} ~ \phi \left( \frac{r_e}{b} \right) \right]^{3/2} \,, 
\end{alignat}
where $\phi$ is determined numerically from Eq.~(\ref{eq:thomas_fermi_eqn}).
With this, Eq.~(\ref{eq:dsigmaStoS}) becomes
\begin{align}
    \label{eq:sigma_s_intermediate}
    \dd \sigma &= \frac{\dd \Omega_e }{(2\pi)^2} \, m_e p_e \left( \frac{2\pi \, \alpha}{3} \right)^2 \, \left| Z \, \psi_{e,f}^*(0) \, \psi_{e,i}(0) - \int \dd ^3 \mathbf{r}_e ~ \psi_{e,f}^* (\mathbf{r}_e) \, \psi_{e,i}(\mathbf{r}_e) \, \big(n_\text{TF}(\mathbf{r}_e) - |\psi_{e,i}(\mathbf{r}_e)|^2 \big) \right|^2
    \nonumber \\
    & \qquad \qquad \qquad \qquad \quad ~~ \times \left| \int \dd ^3 \mathbf{r} ~ \Big(\frac{\mu}{m_N}\Big)^2 r^2 \, \psi_{\x,f}^*(\mathbf{r}) \, \psi_{\x,i}(\mathbf{r}) \right|^2 \, ,
\end{align}
where we have used $r_N = (\mu / m_N) \, r$.

We now turn our attention to obtaining the wavefunctions in Eq.~(\ref{eq:sigma_s_intermediate}), beginning with the initial and final state $\x$ wavefunctions. These wavefunctions are eigenstates of $\hat{T} + \hat{V}_{\x N}$, which are naturally obtained in the $\x$-$N$ center-of-mass frame using the variables defined above: $\mathbf{r} \equiv \mathbf{r}_N - \mathbf{r}_\x$, such that $\mathbf{r}_N = (\mu / m_N) \, \mathbf{r}$ and $\mathbf{r}_\x = -(\mu / m_\x) \, \mathbf{r}$. The initial state $\x$ wavefunction $\psi_{\x,i}$ is an $l = 0$ state with energy $E = k^2 / 2 \mu$ and momentum $k$ (in the $\x$-$N$ frame); based on the decomposition of continuum wavefunctions in a central potential given in Eq.~\eqref{eq:continuum_psi_basis}, we have
\begin{alignat}{1}
    \psi_{\x,i} (r) = \frac{ R_{k0}^\x(r)}{2k \, \sqrt{v_\text{rel}}}  \,,
\end{alignat}
where $v_\text{rel}$ is the velocity of the incoming $\x$ particle in the $\x$-$N$ center-of-mass frame, and $1/\sqrt{v_\text{rel}}$ normalizes $\psi_{\x,i}$ to the ``one particle unit current density" prescription, which is necessary to ensure that the right-hand-side of Eq.~\eqref{eq:fermis_golden_rule} is correctly scaled to give the cross section on the left-hand-side~\cite{Landau1981Quantum}. $R_{k0}^\x(r)$ is the solution to Eq.~\eqref{eq:schrodinger_radial} with $V = V_{\x N}$, $\mu$ the $\x - N$ reduced mass, and $E = k^2 / 2 \mu \simeq 0$ (we neglect the kinetic energy of the incoming $\x$, which is always small relative to the binding energy $E_B$). Asymptotically far away from the nucleus, $R_{k0}^\x$ should tend to the free continuum solution of Eq.~\eqref{eq:schrodinger_radial} (i.e., the solution with $V = 0$ and $l = 0$), which is $2k j_0(kr)$~\cite{Landau1981Quantum}. In the $k \ll 1 / r$ limit, we therefore expect $\lim_{r \to \infty}\psi_{\x,i}(r) = 1/\sqrt{v_\text{rel}}$. To incorporate the effect of the potential $V_{\x N}$ on $\psi_{\x , i}(r)$ at smaller radii, we parametrize the general solution as
\begin{equation}
\psi_{\x,i}(r) = G(r)/\sqrt{v_\text{rel}}
~~,~~ \lim_{r \to \infty} G(r) = 1
~,
\end{equation}
for some dimensionless function $G(r)$ that we compute numerically by solving Eq.~\eqref{eq:schrodinger_radial} for the $\x - N$ system. 

The wavefunction of the final state $\x$, $\psi_{\x,f}$, is a bound eigenstate of the potential $V_{\x N}$ with $l = 0$; we obtain the radial component $R_\text{b.s.}^\x (r)$ of $\psi_{\x,f}$ by again numerically solving Eq.~\eqref{eq:schrodinger_radial} for the $\x-N$ system but with $E = - E_B$ and adjusting the effective coupling $\qeff$ until we obtain a solution that goes to zero as $r \to \infty$ with no nodes. Including the angular piece of the wavefunction (i.e., the spherical harmonic $Y_{00} = 1 / \sqrt{4 \pi}$), we have
\begin{alignat}{1}
    \psi_{\x,f} (r) = R_\text{b.s.}^\x(r) / \sqrt{4\pi} \, .
    \label{eq:chi_yukawa_bound_state_wavefunction}
\end{alignat}

Next, we consider the initial and final state wavefunctions for the ejected electron. From Eq.~(\ref{eq:sigma_s_intermediate}), we see that the capture cross section is enhanced by $Z^2$ for electronic wavefunctions that have non-vanishing weight at the origin. Since wavefunctions of angular momentum $\ell$ scale as $\sim r^\ell$ at small radii, we focus on $s$-wave electronic wavefunctions. The electron is treated as initially occupying a bound eigenstate of $\hat{T} + \hat{V}_{e A}^{(0)}$. For simplicity, we adopt the Roothaan-Hartree-Fock electronic wavefunctions for the atomic orbitals, where the radial part of the wavefunctions is decomposed into a linear combination of Slater orbitals, as computed in Refs.~\cite{Bunge:1993jsz,McLean:1981mjg} (see also Ref.~\cite{xeni_thesis}). This decomposition can be written as 
\begin{alignat}{1}
    R_{nl}^e(r_e) = \sum_j C_{jnl} \, S_{jl}(r_e)
    ~~,~~
    S_{jl}(r_e) = \frac{(2 Z_{jl})^{n_{jl} + 1/2}}{\sqrt{(2 n_{jl})!}} \, a_0^{-3/2} \left(\frac{r_e}{a_0}\right)^{n_{jl} - 1} e^{- Z_{jl} \, r_e / a_0}
    ~,
\end{alignat}
where $C_{jnl}$ is the weight given to each Slater orbital (indexed by $j$), $Z_{jl}$ is an effective charge, $n_{jl}$ is the principal quantum number of that Slater orbital (see, e.g.,  Table~4.1 of Ref.~\cite{xeni_thesis}), and $a_0 = 1 / \alpha m_e$ is the Bohr radius. For a $\x - \text{Xe}$ binding energy of $E_B = \SI{2.5}{\kilo\eV}$, only the $3s$, $4s$, and $5s$ electrons in xenon are shallowly bound enough to be ejected in an $s$-to-$s$ transition. The full initial state wavefunction of the electron for each of these states is then simply
\begin{alignat}{1}
\label{eq:Psiei}
    \psi_{e,i} = \frac{R_{n0}^e(r_e)}{\sqrt{4\pi}}  \, .
\end{alignat}
after including the appropriate spherical harmonic. 

For the outgoing electron, we obtain the radial wavefunction $R_{p_e 0}^e(r_e)$ by numerically solving Eq.~\eqref{eq:schrodinger_radial} for the $e-A$ system, i.e., with $\mu = m_e$, $V = V_{e A}^{(0)}$, $l = 0$, and $E = p_e^2 / 2 m_e = E_B - \omega_{e,i}$. Asymptotically far away from the nucleus, we again expect $R_{p_e 0}^e(r_e)$ to approach the free continuum solution $2 p_e j_0(p_e r)$ up to a phase, which sets the normalization of $R^e_{p_e 0}(r_e)$. The full final electron wavefunction is then (see Eq.~\eqref{eq:continuum_psi_basis} for the normalization factor)
\begin{alignat}{1}
    \psi_{e,f}(r_e) = \frac{ R^e_{p_e 0}(r_e)}{2 p_e} \,. 
\end{alignat}

\begin{figure}
    \centering
    \includegraphics[width=0.6\textwidth]{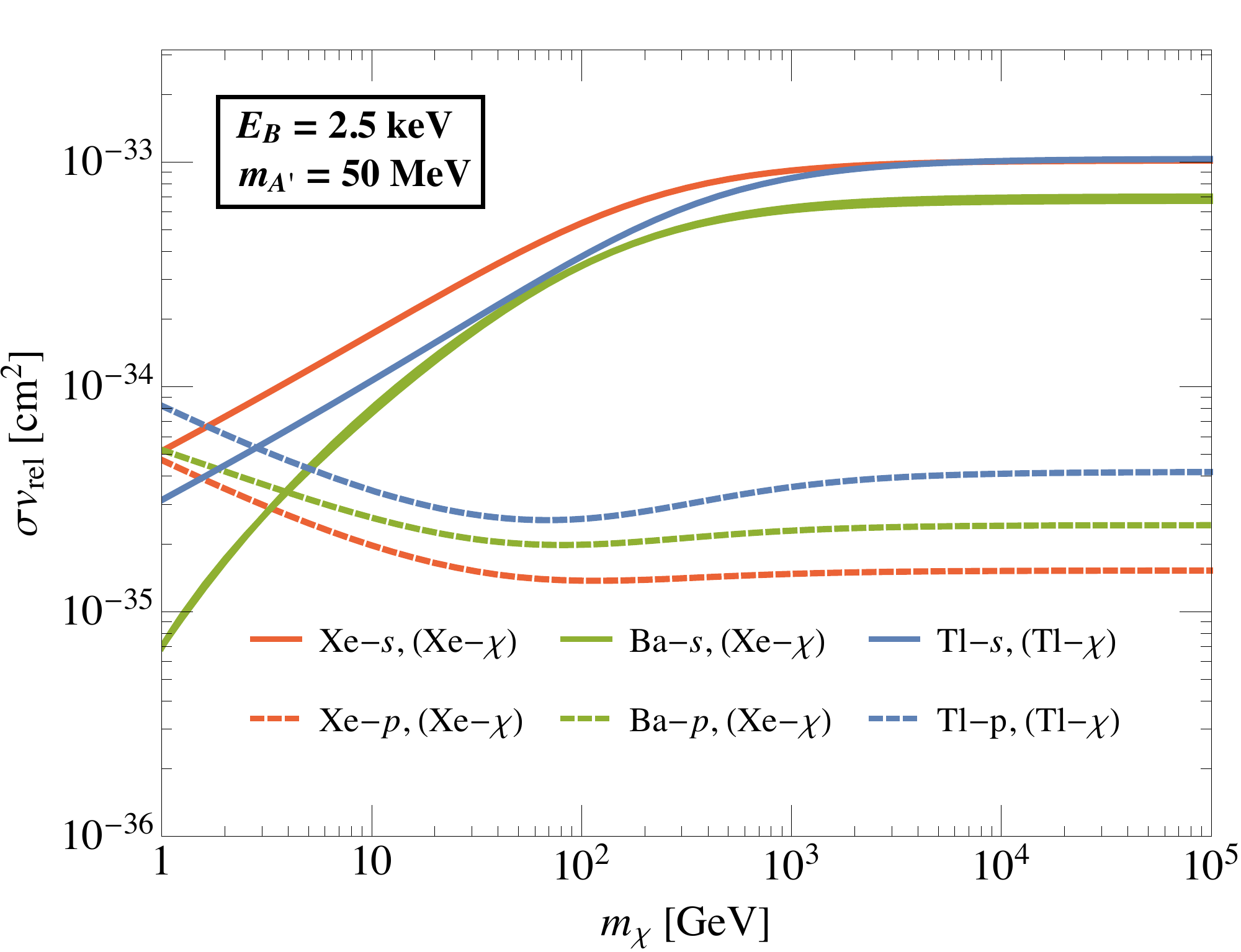}
    \caption{The capture cross sections for the $s$-to-$s$ process (solid) and $p$-to-$s$ process (dashed), for capture of $\x$ onto xenon (red), barium (green), and thallium (blue), as a function of dark matter mass $m_\x$. Cross sections for Xe and Ba assume the existence of a $\chi - \text{Xe}$ bound state  with $E_B = \SI{2.5}{\kilo\eV}$, while the calculation for Tl assumes that $\x$ binds to Tl with $E_B = \SI{2.5}{\kilo\eV}$.}
    \label{fig:capturexsec}
\end{figure}

Having determined the wavefunctions in Eq.~(\ref{eq:sigma_s_intermediate}), we are now ready to compute $\sigma_s$, the $s$-to-$s$ capture cross section after summing over all possible initial electron states. Returning to Eq.~\eqref{eq:sigma_s_intermediate}, we find that the contribution to $V_{e A}^{(2)}$ from the screening electrons (the second term in the first set of vertical brackets) is subdominant (at the level of $\sim 0.5\%$) to the $Z^2$ contribution from the nucleus itself (the first term in the first set of vertical brackets). Keeping only this $Z^2$ contribution yields
\begin{alignat}{1}
    \dd \sigma_s &=2 \, \sum_n \frac{\dd \Omega_e}{(2\pi)^2} \, m_e p_e \, \left(\frac{2 \pi \, Z \alpha}{3} \right)^2 \left(\frac{\mu}{m_N}\right)^4 \, \left| \psi_{e,f}^*(0) \, \frac{R_{n0}^e (0)}{\sqrt{4\pi}} \, \int \dd r ~ 4 \pi r^4 \, \frac{R_\text{b.s.}^\x(r) }{\sqrt{4\pi}} \, \frac{G(r)}{\sqrt{v_\text{rel}}}  \right|^2 \,, 
\end{alignat}
where for $E_B = 2.5 \ \text{keV}$ in xenon the sum is over the $n = $ 3, 4, and 5 electrons, and the factor of two accounts for the pair of electrons in each of these $s$ orbitals. Also note that $p_e = \sqrt{2 m_e (E_B - \omega_{e, i})}$ depends on $n$ through the orbital-dependent electron binding energy $\omega_{e, i}$. Integrating over the outgoing electron solid angle, the above expression simplifies to
\begin{alignat}{1}
    \sigma_s v_\text{rel} = \frac{4\pi\, (Z\alpha)^2}{9} \, \left(\frac{\mu}{m_N}\right)^4  \left| \int \dd r ~ r^4 \, R_\text{b.s.}^\x(r) \, G(r) \right|^2 ~ \sum_n \, 2 \, m_e p_e \, \left| \psi_{e,f}^*(0) \, R_{n0}(0) \right|^2 \,.
\end{alignat}
Noting that the continuum wavefunctions $G$ and $\psi_{e, f}$ are dimensionless, while the bound state wavefunctions $R_\text{b.s.}^\x$ and $R_{n0}$ have dimension $[\text{length}]^{-3/2}$, we can use the characteristic length scale of the $\x$-$N$ bound state $1/\sqrt{2 \mu E_B}$ as well as the size of the atom $1/(Z \alpha m_e)$ to construct the following dimensionless form factors: 
\begin{alignat}{1}
    F_{\x,s}^2 = (2 \mu E_B)^{7/2} \, \left| \int \dd r ~ r^4 \, R_\text{b.s.}^\x (r) \, G(r) \right|^2 
    ~~,~~ 
    F_e^2 = \frac{2}{(Z \alpha m_e)^3} \sum_n \frac{p_e}{m_e} \, \left| \psi^*_{e,f}(0) \,  R_{n0}(0) \right|^2
    ~.
\end{alignat}
Our final result for the cross section is then
\begin{alignat}{1}
    \sigma_s v_\text{rel} &= \frac{4 \pi}{9} \, \frac{(Z \alpha m_e)^5}{(2 \mu E_B)^{7/2}} \, \Big(\frac{\mu}{m_N}\Big)^4 \, F_{\x,s}^2 \, F_e^2 
    \,,
\end{alignat}
or numerically 
\begin{alignat}{1}
    \sigma_s v_\text{rel} &\simeq \SI{7e-34}{\centi\meter\squared} \times \left(\frac{Z}{54}\right)^5 \bigg(\frac{\SI{122}{\giga\eV}}{m_N}\bigg)^4 \bigg(\frac{\SI{2.5}{\kilo\eV}}{E_B}\bigg)^{7/2} \bigg(\frac{\mu}{\SI{100}{\giga\eV}}\bigg)^{1/2} \bigg(\frac{F_{\x,s}^2}{49}\bigg) \bigg(\frac{F_e^2}{0.5}\bigg) \,. 
\end{alignat}
Taking the ionization potentials for the $3s$, $4s$, and $5s$ xenon states to be \SI{1148.7}{\eV}, \SI{213.2}{\eV}, and \SI{23.3}{\eV}~\cite{cardona1978photoemission}, respectively, for $E_B = \SI{2.5}{\kilo\eV}$ we find $F_e^2 = 0.5$ with a relative contribution of 81\%, 17\% and 2\% from each state. For $F_{\x,s}^2$, we find only a weak dependence on $m_\x$, ranging from $F_{\x,s}^2 \approx 35$ for $m_\x = \SI{1}{\giga\eV}$ to $F_{\x,s}^2 \approx 62$ for $m_\x \gg m_\text{Xe}$. A numerical fit to the capture cross section in Xe with $E_B = \SI{2.5}{\kilo\eV}$ gives
\begin{alignat}{1}
    \sigma_{\text{Xe},s} v_\text{rel} \simeq \SI{6e-34}{\centi\meter\squared} \times \left(\frac{\mu}{\SI{100}{\giga\eV}}\right)^{0.55} \,.
\end{alignat}
Fig.~\ref{fig:capturexsec} shows $\sigma_{\text{Xe},s} v_\text{rel}$ as a function of $m_\x$. In addition, we also show \emph{i)} the capture cross section in barium, assuming $\x$ binds to Xe with $E_B = \SI{2.5}{\kilo\eV}$, which determines the survival probability of an incoming $\x$ travelling through the terrestrial overburden of underground detectors, and \emph{(ii)} the capture cross section in thallium, assuming $\x$ binds to Tl with $E_B = \SI{2.5}{\kilo\eV}$, which is relevant for signals in DAMA (see the main body for further discussion). 

\section*{p-to-s Capture Cross Section}

We now consider the case of $p$-to-$s$ capture through $\hat{V}_{e A}^{(1)}$. This process results in a dipole transition of the final state electron, and bears many similarities to the photoelectric cross section of the atom. 

To begin, let us consider the matrix element of Eq.~(\ref{eq:fermis_golden_rule}), which in position space involves $V_{eN}^{(1)} =  \mathbf{r}_N^i \, \partial^i \, V_{e A}^{(0)}$. In the language of operators, this corresponds to $\hat{V}_{eN}^{(1)} = i \, \hat{\mathbf{r}}_N^i \, \, [ \, \hat{\mathbf{p}}_e^i, \hat{V}_{e A}^{(0)} \, ]$. The matrix element of $\hat{V}_{eN}^{(1)}$ between the initial state $|i \rangle = | \chi_i \rangle \, | e_i \rangle$ and final state $|f \rangle = | \chi_f \rangle \, | e_f \rangle$ is then
\begin{equation}
\langle f | \hat{V}_{eN}^{(1)} | i \rangle = i \, \langle \chi_f | \hat{\mathbf{r}}_N^i | \chi_i \rangle ~ \langle e_f | \,  [ \, \hat{\mathbf{p}}_e^i, \hat{V}_{e A}^{(0)} \, ] \, | e_i \rangle
= i \, \langle \chi_f | \hat{\mathbf{r}}_N^i | \chi_i \rangle ~ \langle e_f | \,  [ \, \hat{\mathbf{p}}_e^i, \hat{H}^{(0)} \, ] \, | e_i \rangle
= - i E_B \, \langle \chi_f | \hat{\mathbf{r}}_N^i | \chi_i \rangle ~ \langle e_f | \hat{\mathbf{p}}_e^i | e_i \rangle
~,
\end{equation}
where in the second and third equalities we have made use of the fact that the only contribution to $\hat{H}_0$ that does not commute with $\hat{\mathbf{p}}_e$ is $\hat{V}_{eA}$, and that the initial and final electron states are both eigenstates of $\hat{H}_0$, with the difference in energy between the two states being given by the $\x$-$N$ binding energy, $E_B$. The square of this matrix element can be simplified by averaging over the direction of the electron momentum and using Eq.~(\ref{eq:components_in_integral_identity}), which allows us to replace 
\begin{equation}
\label{eq:VeNmatrix1}
\big|\langle f | \hat{V}_{eN}^{(1)} | i \rangle \big|^2 \to \frac{1}{3} \, E_B^2 \, \big| \langle \chi_f | \hat{\mathbf{r}}_N | \chi_i \rangle \big|^2 ~ \big| \langle e_f | \hat{\mathbf{p}}_e | e_i \rangle \big|^2
~.
\end{equation}
Notice that we have successfully factored the contributions into a DM-only piece and an electron-only piece. The electron part of the matrix element is the same one that appears in electromagnetic dipole transitions. In particular, the photoelectric cross section in the long wavelength limit (averaging over photon polarization) is $ \sigma_\text{pe} = 2 \alpha p_e \, \big| \langle e_f | \hat{\mathbf{p}}_e | e_i \rangle \big|^2 / (3 m_e E_B)$~\cite{Sakurai:2011zz} for an incoming photon with energy $E_B$ and outgoing electron with momentum $p_e$. This allows Eq.~(\ref{eq:VeNmatrix1}) to be rewritten as
\begin{equation}
\label{eq:VeNmatrix2}
\big|\langle f | \hat{V}_{eN}^{(1)} | i \rangle \big|^2 \to  ~ \frac{m_e E_B^3}{2 \alpha p_e} \, \sigma_\text{pe} ~ \big| \langle \chi_f | \hat{\mathbf{r}}_N | \chi_i \rangle \big|^2 
~.
\end{equation}

The capture cross section is determined from this matrix element as in Eq.~(\ref{eq:fermis_golden_rule}). Rewriting the DM matrix element in position basis and using $r_N = (\mu / m_N) \, r$, we have
\begin{alignat}{1}
\label{eq:sigmapwave1}
    \dd \sigma_p = \frac{\dd \Omega_e}{(2\pi)^2} \,  \frac{m_e^2 E_B^3}{2 \alpha} \,  \left(\frac{\mu}{m_N}\right)^2 \, \sigma_\text{pe} \, ~  \left| \int \dd^3 \mathbf{r} ~ \psi_{\x,f}^*(\mathbf{r}) \,  \psi_{\x,i}(\mathbf{r}) ~ \mathbf{r} \right|^2 \,.
\end{alignat}
The DM wavefunctions can be derived in a similar manner to the  the previous section. The final bound state wavefunction $\psi_{\chi, f}$ is identically given by the $s$-wave state in  Eq.~\eqref{eq:chi_yukawa_bound_state_wavefunction}. Since $\psi_{\chi, f}$ is spherically symmetric, we see that the integral in Eq.~(\ref{eq:sigmapwave1}) vanishes if $\psi_{\chi, i}$ is also $s$-wave. Hence, we take the incoming wavefunction $\psi_{\chi , i}$ as the $\ell = 1$ term in the continuum wavefunction expansion in Eq.~\eqref{eq:continuum_psi_basis}, which gives
\begin{alignat}{1}
    \psi_{\x,i} = \frac{3 i}{2k \sqrt{v_\text{rel}}} \,  (\hat{\mathbf{k}} \cdot \hat{\mathbf{r}}) \, R_{k1}(r) \,,
\end{alignat}
where $R_{k1}(r)$ is the solution to Eq.~\eqref{eq:schrodinger_radial} with $l = 1$, $V = V_{\x N}$, and $E \simeq 0$. Far away from the nucleus, $R_{k1}$ should tend toward the free $l = 1$ solution $ R_{k1} (r) \to 2 k j_1(k r) \simeq 2 k^2 r / 3$ for $k \ll 1/r$. As in the last section, to incorporate the effect of the potential $V_{\x N}$ on $\psi_{\x , i}(r)$ at smaller radii, we parametrize the general solution as
\begin{alignat}{1}
    \psi_{\x,i} = i (\mathbf{k} \cdot \mathbf{r}) \, F(r)  / \sqrt{v_\text{rel}}
    ~~,~~ \lim_{r \to \infty} F(r) = 1
    ~,
\end{alignat}
for some dimensionless function $F(r)$ that we compute numerically by solving Eq.~\eqref{eq:schrodinger_radial} for the $\x - N$ system. 

Having determined the DM wavefunctions, Eq.~(\ref{eq:sigmapwave1}) reduces to
\begin{alignat}{1}
    \sigma_p v_\text{rel} = \frac{2 m_e^2 \, E_B^3 \, \mu^3 \, T}{3 \alpha \, m_N^2} \, \sigma_\text{pe} \, \left| \int \dd r ~ R_\text{b.s.}^\x(r) \, F(r) \, r^4 \right|^2 \, ,
\end{alignat}
where we integrated over the outgoing electron solid angle and used Eq.~(\ref{eq:components_in_integral_identity}) to replace $(\mathbf{k} \cdot \mathbf{r}) \, \mathbf{r} \to (r^2 / 3) \, \mathbf{k}$ in the integral over $r$. Finally, we replaced $k^2$ with its thermally-averaged value $k^2 \to 3 \mu T$, where $T \simeq 300 \ \text{K}$ is the temperature of the thermalized $\chi$ particle; as a result, the cross section is suppressed by the small thermal velocity. As in the last section, we can construct the following dimensionless quantity
\begin{alignat}{1}
    F_{\x,p}^2 = (2 \mu E_B)^{7/2} \left| \int \dd r ~ R_\text{b.s.}^\x(r) \, F(r) \, r^4 \right|^2 \,,
\end{alignat}
such that the $p$-to-$s$ capture cross section is given by
\begin{alignat}{1}
    \sigma_p v_\text{rel} = \frac{m_e^2 \, T \, F_{\x,p}^2 \, \sigma_\text{pe}}{12 \alpha \, m_N^2 \, (2 \mu E_B)^{1/2}} ~.
\end{alignat}
For parameters that are representative of $E_B = 2.5 \ \text{keV}$ in xenon, we find
\begin{alignat}{1}
    \sigma_p v_\text{rel} \simeq \SI{e-35}{\centi\meter\squared} \times \bigg(\frac{T}{\SI{300}{\kelvin}}\bigg) \bigg(\frac{\SI{122}{\giga\eV}}{m_N}\bigg)^2 \bigg(\frac{\SI{2.5}{\kilo\eV}}{E_B}\bigg)^{1/2} \bigg(\frac{\SI{100}{\giga\eV}}{\mu}\bigg)^{1/2} \bigg(\frac{F_{\x,p}^2}{176}\bigg) \bigg(\frac{\sigma_{pe}}{\SI{2.5e-19}{\centi\meter\squared}}\bigg) \,.
\end{alignat}
The photoelectric cross section $\sigma_\text{pe}$  for various elements, including xenon, with a photon energy of \SI{2.5}{\kilo\eV} can be interpolated from data in Refs.~\cite{West:1978pto,Veigele:1973tza}. Fixing $E_B = \SI{2.5}{\kilo\eV}$ in xenon, we find only a weak dependence of $F_{\x,p}^2$ on $m_\x$, ranging from $F_{\x,p}^2 \approx 81$ for $m_\x = \SI{1}{\giga\eV}$ to $F_{\x,p}^2 \approx 287$ for $m_\x \gg m_N$. 

The dashed lines in Fig.~\ref{fig:capturexsec} show $\sigma_p v_\text{rel}$ for various elements as a function of $m_\x$. For xenon, we find that the $p$-to-$s$ rate is smaller than the $s$-to-$s$ rate, with more than an order of magnitude suppression once $m_\x \gtrsim \SI{10}{\giga\eV}$; for simplicity, we can therefore neglect the contribution of $\sigma_p$ to the full capture cross section in xenon. For capture in barium and thallium, however, the $p$-to-$s$ cross section dominates over the $s$-to-$s$ rate for $m_\x \lesssim \SI{5}{\giga\eV}$.

\end{document}